\documentclass[preprint,5p,times,twocolumn]{elsarticle}

\usepackage[latin1]{inputenc}                    
\usepackage{graphicx}                            
\usepackage{latexsym}                            
\usepackage{amsfonts}                            
\usepackage{amssymb}                             
\usepackage{amsmath}                             
\usepackage[mathscr]{eucal}                      
\usepackage{dcolumn}                             
\usepackage{theorem}                             
\usepackage{footnote}
\usepackage{enumitem}
\usepackage{bm}
\usepackage{color}
\usepackage{natbib}
\usepackage{nccmath}
\usepackage{bm}
\usepackage{slashed}
\usepackage{hyperref}
\usepackage{braket}
\usepackage{float}
\usepackage{lipsum,multicol}
\usepackage{dblfloatfix}

\numberwithin{equation}{section}

\def\be{\begin{equation}}
\def\ee{\end{equation}}
\def\bea{\begin{eqnarray}}
\def\eea{\end{eqnarray}}

\journal{Physics Letters B}

\begin{document}

\title{Effect of the pion field on the distributions of pressure and shear in the proton}
\author{Shiryo~Owa${}^{a}$,
A.~W.~Thomas${}^{a}$,
X.G.~Wang${}^{a}$\\[5mm]
\itshape{$^a$ ARC Centre of Excellence for Dark Matter Particle Physics and CSSM, \\
Department of Physics, University of Adelaide, Adelaide SA 5005, Australia}
}

\date{\today}

\begin{abstract}
In light of recent experimental progress in determining the pressure and shear distributions in the proton, these quantities are calculated in a model with confined quarks supplemented by the pion field required by chiral symmetry. The incorporation of the pion contributions is shown to account for the long-range distributions, in general agreement with the experimentally extracted quark contributions. The results of the model  are also compared with lattice QCD results at unphysically large quark mass.
\end{abstract}


\maketitle

\section{Introduction}
A quantitative understanding of the nucleon structure in terms of quarks and gluons is one of the most challenging topics in modern nuclear physics.
Many aspects of this structure are encoded in form factors, which means that their study is of vital importance to our understanding of QCD.
The gravitational form factors (GFFs), which contain the basic mechanical properties such as mass, angular momentum and internal forces, are currently receiving a great deal of attention~\cite{Polyakov:2018exb,Granados:2019zjw,Chakrabarti:2020kdc,Ji:2021mtz,Fiore:2021wuj,Mamo:2021krl,Freese:2019bhb}.
There are three GFFs, which are related to the nucleon matrix element of the energy moment tensor (EMT)~\cite{Kobzarev:1962wt, Pagels:1966zza}.
The critical $D$-term form factors, which parametrize the spatial components of the EMT, 
carry information about the pressure and shear distributions experienced inside the 
nucleon~\cite{Polyakov:2002yz, Polyakov:2018zvc}.

Theoretical developments connect the GFFs to the moments of generalized parton distributions (GPDs)~\cite{Ji:1996ek, Ji:1996nm}, which are accessible in hard exclusive processes, 
particularly deeply virtual Compton scattering (DVCS).
Recently the quark $D$-term form factor of the proton at a scale $\mu^2 = 1.39\ {\rm GeV}^2$ was extracted from a DVCS experiment at the Thomas Jefferson National Accelerator Facility 
(JLab)~\cite{Girod:2007aa, Jo:2015ema}. This led to the first experimental determination of the pressure distribution and shear forces inside the proton~\cite{Burkert:2018bqq, Burkert:2021ith} (referred to as BEG).

It is worth noting that, in order to extract these distributions, it was  necessary to choose some functional form, in this case a tripole, to extend the form factor, $D(t)$, beyond the limited range of momentum transfer accessible in the experiment. This does introduce some degree of model dependence into the results. This same $D$-term form factor was also used to derive the 2-dimensional transverse light front pressure density~\cite{Freese:2021qtb}.
Most recently, the first attempt to extract experimental information on the gluon $D$-term form factor was carried out using the photoproduction of the $\phi$ and $J/\psi$ vector mesons near their respective thresholds~\cite{Kou:2021qdc}.
While the individual quark and gluon GFFs depend on the scale $\mu$, the total $D$-term $D_{\rm tot} = D_q + D_g$ is scale invariant as a consequence of conservation of the EMT. 

On the other hand, the GFFs of hadrons can also be computed from lattice QCD (LQCD). 
While the quark GFFs of the proton have been studied extensively~\cite{Hagler:2003jd, Gockeler:2003jfa, Hagler:2007xi, Bratt:2010jn, Alexandrou:2013joa}, 
the first LQCD determination of gluon GFFs was presented very recently~\cite{Shanahan:2018pib}, albeit at a quark mass considerably higher than that found in nature, corresponding to $m_{\pi} = 0.45\ {\rm GeV}$. 
The pressure and shear distributions of the proton were then extracted from the calculated $D$-term form factors~\cite{Shanahan:2018nnv}, finding that the gluon contributions are dominant compared with the quark terms at the scale $\mu^2 = 4\ {\rm GeV}^2$. Once again there is some model dependence in these results associated with the need to assume a functional form for $D(t)$.

There have also been a number of theoretical studies of the proton's GFFs. 
The MIT bag model, which is attractive due to its simplicity and incorporation of confinement, was the first to be applied to GPDs and EMT form factors~\cite{Ji:1997gm}.
This framework was recently extended to study the pressure and shear distributions inside the 
proton~\cite{Neubelt:2019sou}. 
The first chiral calculation of the pressure and shear in the nucleon was carried out some time ago 
by Goeke and collaborators~\cite{Goeke:2007fp}, using the chiral quark soliton model. For a recent formal study of the chiral corrections to the EMT we refer to Ref.~\cite{Alharazin:2020yjv}.

Here we calculate the pressure distribution and shear forces in the proton using the cloudy bag model (CBM), in which the nucleon is treated as a three quark bag surrounded by a perturbative pion 
cloud~\cite{Theberge:1980ye, Thomas:1981vc, Theberge:1981pu, Thomas:1982kv}. 
Since Neubelt {\em et al.}~\cite{Neubelt:2019sou} have recently calculated the valence quark contribution, 
we focus on the contribution generated by the pion cloud. The pionic contribution to the quark and gluon 
$D$-terms in the EMT of the nucleon will be shown to be a product of the distribution of pions in the nucleon, 
$D_{\pi/N}(t)$, which is scale independent, with the quark and gluon $D$-terms of the pion, $D^\pi_{q,g}$. 
For the latter we use the LQCD calculations of Brommel~\cite{Brommel:2007zz} and 
Pefkou {\em et al.}~\cite{Pefkou:2021fni}, respectively.

\section{Theoretical framework}
Within the CBM the dressed nucleon state vector, $\ket{\,\tilde{N}}$, may be expressed as 
\be
\ket{\,\tilde{N}} \, = \sqrt{Z} \left( \, \ket{\,N}+\ket{\,N\pi}+\ket{\,\Delta\pi}+{\rm higher \, order \,  terms}
 \right), 
\label{eq:N}
\ee
where the $\ket{\,N\pi}$ and $\ket{\,\Delta\pi}$ components have probabilities of order 20-25\% and 10\%, respectively~\cite{Thomas:2001kw}, and the higher order terms are small~\cite{Dodd:1981ve}. There have been very detailed studies of the structure of the nucleon within the CBM. The model provides a good description of nucleon properties from electromagnetic form factors~\cite{Theberge:1981pu} to the parton distribution functions~\cite{Schreiber:1991qx}, including nucleon spin~\cite{Schreiber:1988uw,Thomas:2008ga}.

The coupling of the pion to the three-quark bag is
\be
H_{int} \, = \, \sum_{B,B'} \, \int d^3\vec{k} (B^\dagger \, B' \, a_{\vec{k}} \, v_{\vec{k}}^{BB'} \, + \, h.c.) \, ,
\label{eq:Hint}
\ee
where $h.c.$ denotes the Hermitian conjugate. $B^\dagger$ and $B$ respectively create and destroy the appropriate three-quark bag. For the nucleon, for example, the vertex function, which suppresses the coupling of high momentum pions because of the  extended size of the nucleon bag, is calculated directly from the model
\be
v_{\vec{k},a}^{NN} \, = \, \frac{i}{\sqrt{(2\pi)^3 \omega_k}} \, \frac{g_A}{2f_\pi} \, u(k) \,  \tau_a \, \vec{\sigma} . \vec{k} \, .
\label{eq:vertex}
\ee
Here $\omega_k = \sqrt{k^2 + m_\pi^2}$, $g_A$ is the axial charge of the nucleon (1.27), $f_\pi$ is the pion decay constant (93 MeV) and $u(k)$ is the form factor, 
$u(k) = 3 j_1(kR)/kR$, calculated in the CBM, where $R=0.8$ fm is the radius of the bag.

Denoting the contribution to the EMT of the nucleon from the pion in the $\ket{\,N\pi}$ component of the nucleon wave function as $\bra{\, N\,}  T^{ij}(0) \ket{\,N\,}_{\pi N}$, we find (in the Breit frame for momentum transfer $\vec{\Delta}$)
\begin{align}
	&\bra{\, N( \vec{\Delta}/2),s\,} T^{ij}(0) \ket{\, N( -\vec{\Delta}/2),s\,}_{\pi N} \nonumber \\ 
	&=\, \frac{3 g_A^2}{32 \pi^3 f_\pi^2} \, \int d^3\vec{k} \, \frac{ E_{|\vec{k}+\vec{\Delta}/2|}\, u(k)\, u(\,|\vec{k}+\vec{\Delta}|\,) \, (k^2+\vec{k} \cdot \vec{\Delta})}{2 \omega_k^2 \, \omega_{|\vec{k}+\vec{\Delta}|}^2} \nonumber \\ 
	&\mkern20mu \bra{\,\pi(\vec {k}+\vec{\Delta})\,}  T^{ij}(0) \ket{\,\pi(\vec{k})\,}. 
\label{eq:start}
\end{align}
Here, as usual, $E_q$ and $\omega_q$ denote the energy of a nucleon or pion of momentum $\vec{q}$, $s$ is the nucleon spin and following Ref.~\cite{Pefkou:2021fni} all hadron states are normalised as 
$\braket{\,\vec{p}\,|\,\vec{p}\,^\prime} = (2\pi)^3 \, 2p^0 \, \delta(\vec{p} - \vec{p}\,^\prime)$. 

The $\Delta$-nucleon coupling takes a similar form, namely, in Eq.~(\ref{eq:vertex}) we replace $\vec{\sigma}$ by $\vec{S}$ the transition spin operator and $\tau_a$ by $T_a$ the transition isospin operator, and insert the appropriate SU(6) 
coefficient~\cite{Theberge:1980ye}. Furthermore, the energy denominators $\omega_q^2$ are replaced with $\omega_q (\omega_q + \Delta M)$ in Eq.~(\ref{eq:start}), where $\Delta M$ is the 300 MeV mass difference of the nucleon and $\Delta$.

Using the standard expressions for the EMT of the nucleon and pion, for example in the pion case
\bea
&& \bra{\,\pi(\vec{\Delta}/2)\,} T_{q,g}^{ij}(0) \ket{\,\pi(-\vec{\Delta}/2)\,} \nonumber \\
&=& 2K^i K^j A_{q,g}(t) \, + \, 2 m_\pi^2 \bar{c}_{q,g}^\pi(t) \, +\frac{1}{2} (\Delta^i \Delta^j 
- g^{ij} \Delta^2)D_{q,g}^\pi(t) \, , \nonumber \\
\label{eq:EMTpi}
\eea
where $t=\Delta^2$ ($t=-\vec{\Delta}^2$ in the Breit frame) and the subscripts denote the quark and gluon components of the respective EMTs.
We then see that the pion contribution to the $D$-term of the nucleon associated with the $\ket{\,N\pi}$ component of the wave function $D^{\pi N}$ is directly related to the corresponding $D$-term of the pion by
\be
D_{q,g}^{\pi N}(t) \, = \, D_{\pi/N}(t) \, D_{q,g}^\pi(t) \, ,
\label{eq:convolution}
\ee
where 
\bea
D_{\pi/N} &=& \frac{3 g_A^2}{32 \pi^2 f_\pi^2} \int_0^\infty dk \, k^2 \, \frac{u(k)}{k^2+m_\pi^2} 
\nonumber \\
&& \int_{-1}^{+1}dx \frac{u(y) \sqrt{m_N^2 +z}\ (k^2 + k \Delta x)}{y^2+m_\pi^2} \, .
\label{eq:DpiN}
\eea
To simplify the expression in Eq.~(\ref{eq:DpiN}) we have defined $y = \sqrt{k^2+\Delta^2 + 2k \Delta x}$ and $z= k^2 +\Delta^2/4 + k \Delta x$. In addition, there is also the $D_{\pi/\Delta}$ term from the 
$\ket{\, \Delta\pi}$ component, which takes a form similar to Eq.~(\ref{eq:DpiN}), with changes previously described.

The product form of $D^{\pi N}_{q,g}$ given in Eq.~(\ref{eq:convolution}) means that in coordinate space, the $D$-term is a convolution of the distribution of pions in the nucleon with the distribution of quarks or gluons in the pion. As a consequence of the long-range nature of the pion cloud of the nucleon, this means that the pion contribution to the EMT of the nucleon dominates at larger radii. In terms of the Fourier transform of $D_{q,g}^{\pi N}$ 
\be
\widetilde{D}_{q,g}(r) \, = \, 
\int\frac{d^3\vec{\Delta}}{2E_{\Delta /2}(2\pi)^3}e^{-i\vec{\Delta}\cdot\vec{r}}D_{q,g}(-\vec{\Delta}\,^2) \, ,
\label{eq:FT}
\ee
we can readily calculate the pressure and shear distributions~\cite{Polyakov:2002yz,Polyakov:2018zvc}
\begin{align}
	&s(r) = -\frac{1}{2}r\frac{d}{dr}\frac{1}{r}\frac{d}{dr}\widetilde{D}(r) \, , \\
	&p(r) = \frac{1}{3}\frac{1}{r^2}\frac{d}{dr}r^2\frac{d}{dr}\widetilde{D}(r) \, .
\end{align}
%

The pressure and shear distributions for the bare nucleon component $\ket{\, N}$ (the three-quark bag) are calculated using the MIT bag model, under 
the approximation that the 
bag is a static spherical cavity~\cite{DeGrand:1975cf} of radius $R$.
The solution for the ground state quark wave function is
\begin{equation}
\label{eq:quark-wf}
q(\vec{r}) = \frac{N}{\sqrt{4\pi}}
               \left( \begin{array}{c}
                    \alpha_{+} j_0(\frac{\omega r}{R}) \\
                    \alpha_{-} i \vec{\sigma} \cdot \hat{r} j_1(\frac{\omega r}{R})
                         \end{array} 
                \right) \chi_s , 
\end{equation}
where $\alpha_\pm = \sqrt{1\pm m/E}$, $m$ the current quark mass, $\chi_s$ the two-component Pauli spinors, $j_0$ and $j_1$ are the spherical Bessel functions. $E$ is the single quark energy and 
the eigenfrequency $\omega$ is derived from imposition of the linear boundary condition in the usual way.

The spatial components of the symmetric energy-momentum tensor for the valence quarks are
\be
\label{eq:Tij}
T^{ij}_q(\vec{r}) = \frac{N_qN^2}{4\pi}\alpha_+\alpha_-\left[ \left( j_0j'_1-j'_0j_1-\frac{j_0j_1}{r} \right)  \frac{r^i r^j}{r^2} +\frac{j_0j_1}{r}\delta^{ij} \right]\theta_V .
\ee
Here $N_q$ is the number of quarks ($N_q=3$ for a nucleon), and the primes denote 
differentiation with respect to $r$. 
From this it is straightforward to extract the pressure distribution and shear forces in the proton,
\begin{subequations}
\bea
	p(r) &=& \left[ \frac{N_qN^2}{12\pi}\alpha_+\alpha_-\left( j_0j'_1-j'_0j_1+\frac{2j_0j_1}{r} \right)-B \right]\theta_V ,\mkern40mu\\
	s(r) &=& \left[ \frac{N_qN^2}{4\pi}\alpha_+\alpha_-\left( j_0j'_1-j'_0j_1-\frac{j_0j_1}{r} \right) \right]\theta_V .
\eea
\end{subequations}
These quark contributions inside the bag are in agreement with the calculations reported in
 Ref.~\cite{Neubelt:2019sou}.

\section{Results}
The pionic contributions to the $D$-term of the nucleon are directly computed from the product of $D_{\pi/B}$ (where $B$ is either $N$ or $\Delta$) and $D^\pi_{q,g}$. The latter are obtained from LQCD calculations, with the more recent calculations at $m_\pi = 450$ MeV~\cite{Pefkou:2021fni} and the much older pioneering 
work~\cite{Brommel:2007zz}, which has a significantly larger uncertainty, giving the value at the physical mass through an extrapolation formula. We stress that the present calculation of the pion distributions in the nucleon, $D_{\pi/B}$, may be combined with any future calculations of the quark and gluon distributions in the pion when they are generated. To that end we provide a fit to the functions $D_{\pi /N}$ and $D_{\pi /\Delta}$~\cite{dpiBfit}.

The individual quark and gluon EMTs are dependent on renormalisation scale, with the evolution of the operators being the same as the second moments of the parton distribution functions. The scale associated with a quark model like the MIT bag is low, typically~\cite{Schreiber:1991qx} of order 0.2-0.3 GeV$^2$. Therefore, the bag EMT is evolved to $\mu^2 = 1.4\ \mathrm{GeV}^2$ and $\mu^2 = 4\ \mathrm{GeV}^2$ to compare with the experimental data and LQCD calculations, respectively. 
Given the computational complexity of our results, the calculated $D^{\pi B}_{q,g}$ terms are fitted with the usual functional form, namely, the tripole ansatz to the quark and gluon $D$-terms following from \cite{Burkert:2018bqq,Pefkou:2021fni}.
Furthermore, with limited knowledge of the functional form for lattice $D$-terms, we use tripole forms for both quark~\cite{Brommel:2007zz} and 
gluon~\cite{Pefkou:2021fni} components. 

\begin{figure}[t]
	\centering
	\includegraphics[width=\columnwidth]{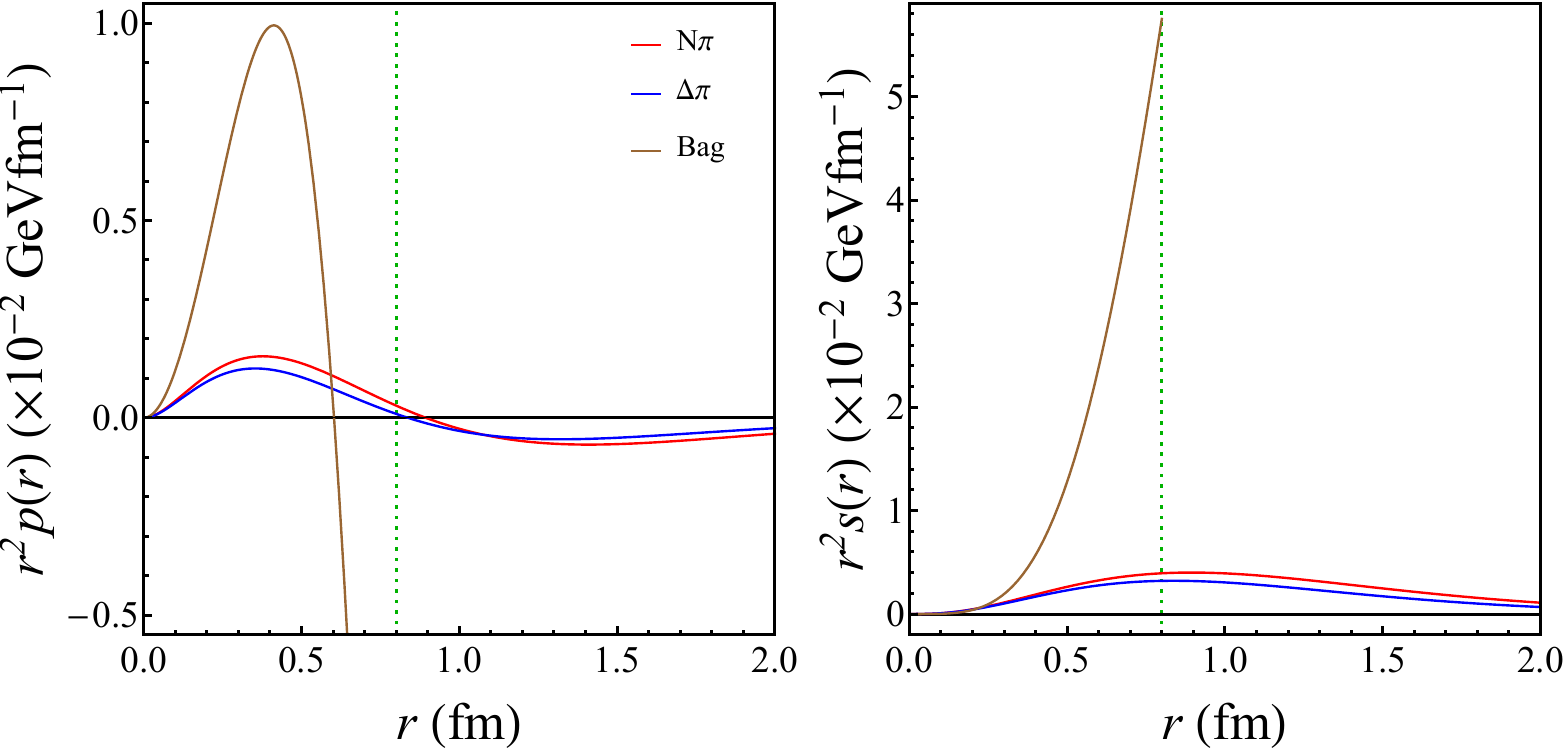}
	\caption{(colour online). The quark pressure and shear distributions inside the proton from the contributions $\ket{\, N}$, $\ket{\, N\pi}$, and $\ket{\, \Delta\pi}$, at quark masses corresponding to the physical pion mass and $\mu^2= 1.4\ \mathrm{GeV}^2$. The green vertical line indicates the bag radius.}
	\label{fig:mpi-phy}
\end{figure}

\begin{figure}[t]
	\centering
	\includegraphics[width=\columnwidth]{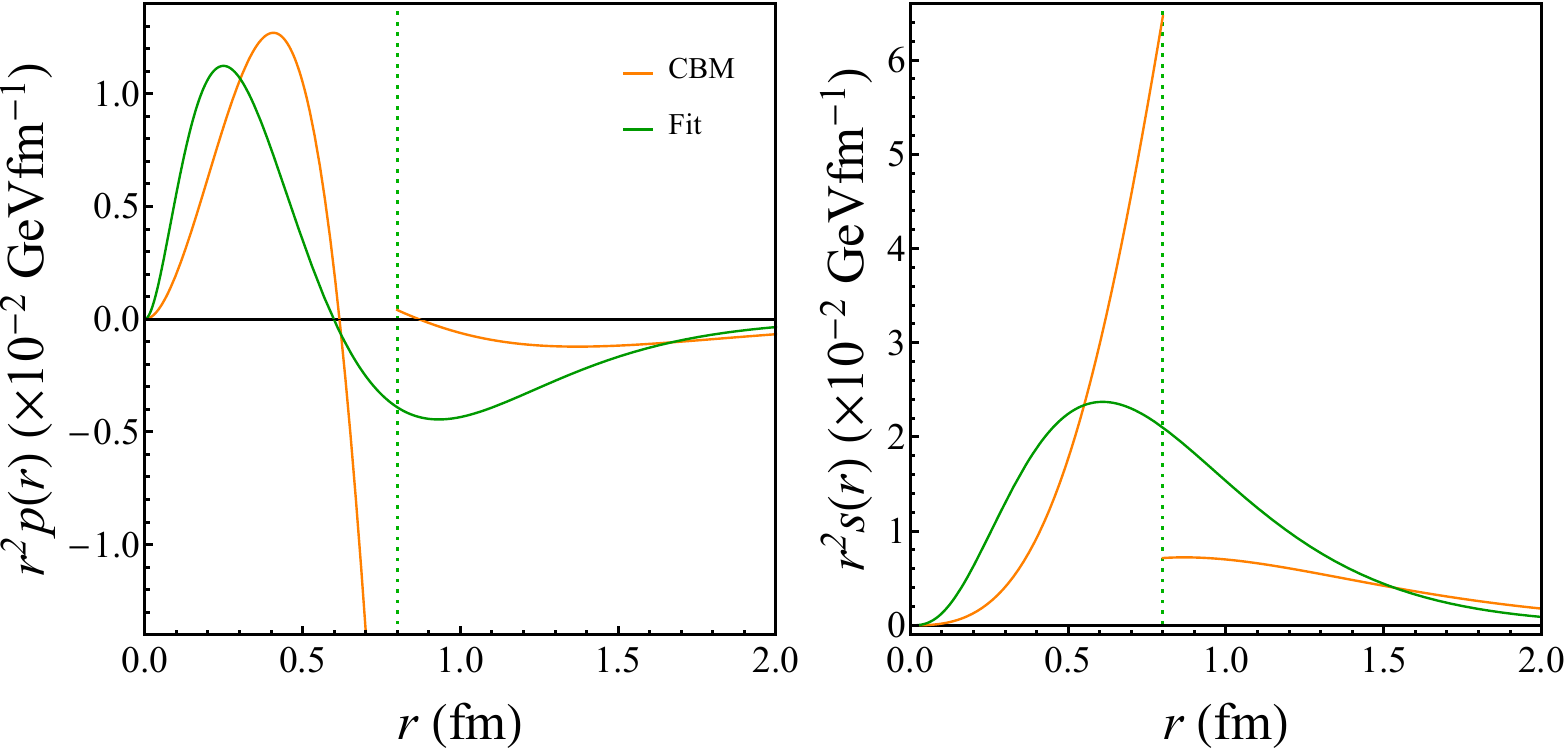}
	\caption{(colour online). Calculated total pressure and shear distributions and the corresponding smooth fits using Eq.~(\ref{eq:Tij}) at $m_\pi = 140$ MeV. The green vertical line indicates the bag radius.}
	\label{fig:mpi-unphy-fits}
\end{figure}

In Fig.~\ref{fig:mpi-phy}, 
we show the individual quark contributions to the pressure distribution and shear in the proton at a quark mass corresponding to the physical pion mass.  In a more sophisticated treatment of confinement one would expect a smooth distribution, rather than the discontinuous behaviour shown in Fig.~\ref{fig:mpi-phy}.  With that in mind, in Fig.~\ref{fig:mpi-unphy-fits} we show a fit to the  distributions at physical pion mass with a smooth function. The fit functions are applied to total pressure and total shear separately, using the simple forms
\be
\label{eq:Tij}
	p(r) = A(1+B\,r)\,e^{-Cr}\ \text{and }\ s(r) = D\,r\,e^{-Er}\, ,
\ee
where $A,B,C,D,$ and $E$ are real constants~\cite{psfits}. 

\begin{figure}[t]
	\centering
	\includegraphics[width=\columnwidth]{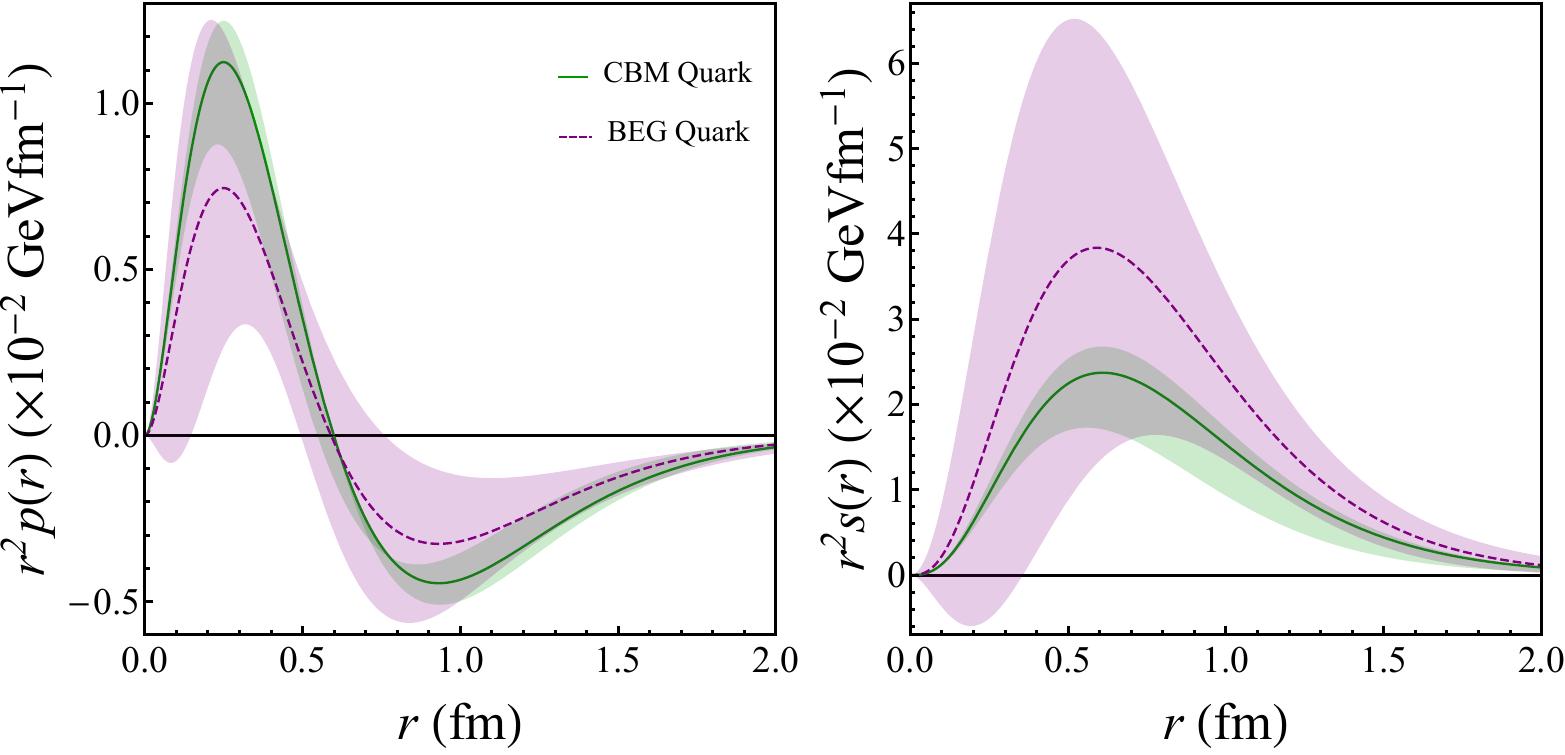}
	\caption{(colour online). The quark pressure and shear distributions fit in the model, evaluated at the physical pion mass and scale $\mu^2=1.4 \rm{\ GeV}^2$, are shown in comparison with the experimental results of Burkert {\it et al.}~\cite{Burkert:2018bqq,Burkert:2021ith}.}
	\label{fig:compareBEG}
\end{figure}

The only experimental data that we have at the present time is the distribution of quark pressure and shear coming from the analysis of Burkert {\it et al.}~\cite{Burkert:2018bqq,Burkert:2021ith}. This is shown in Fig.~\ref{fig:compareBEG} in comparison with the smooth fit to our model calculation. Even allowing for the considerable uncertainties associated with the LQCD calculations of the $D$ function of the pion, it is clear that the calculation reproduces the features of the data quite well, with the contribution from the pion cloud dominating beyond 0.8 fm.

\begin{figure}[t]
	\centering
	\includegraphics[width=\columnwidth]{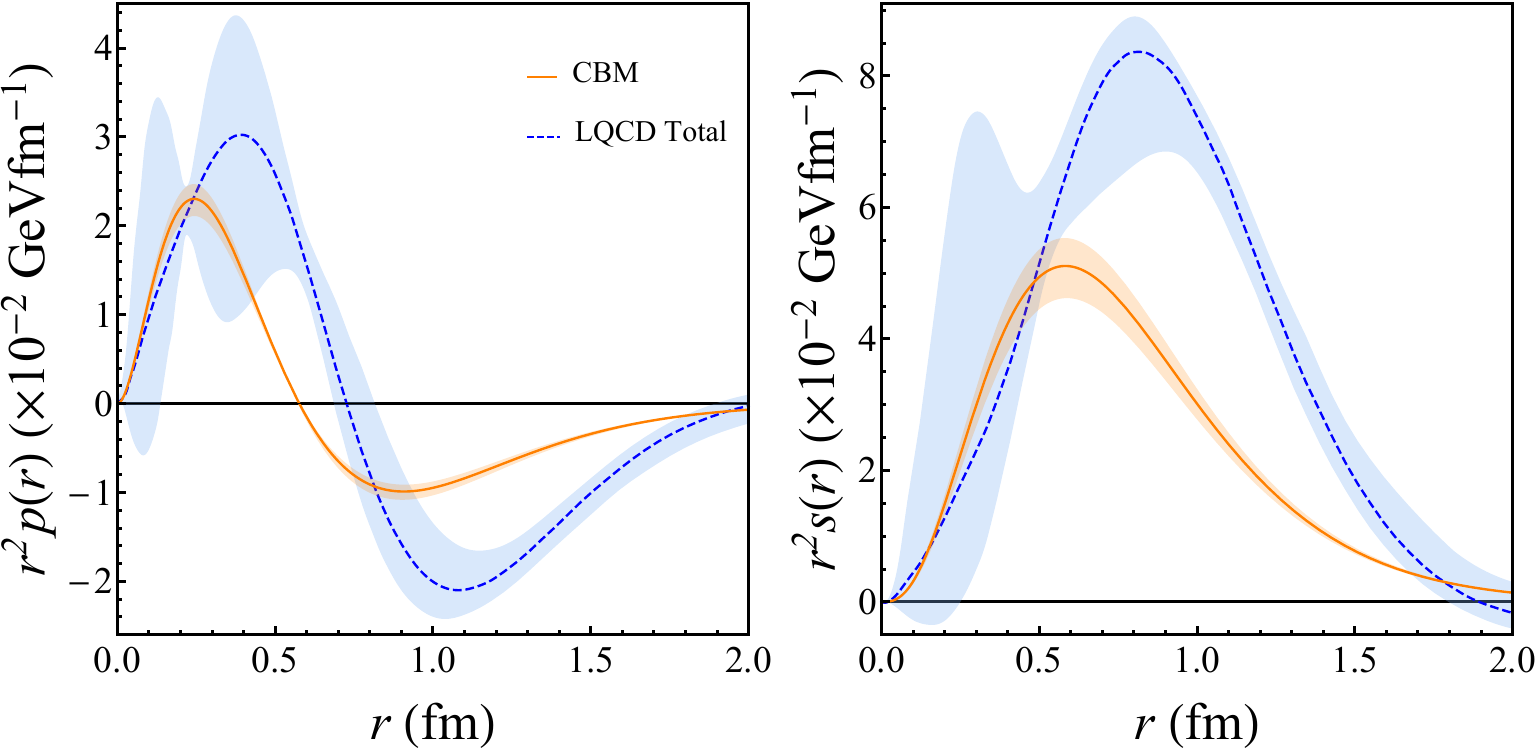}
	\caption{(colour online). The pressure and shear distributions for $m_{\pi} = 450\ {\rm MeV}$ with corresponding quark mass in the bag, in comparison with recent LQCD results~\cite{Shanahan:2018nnv}.}
	\label{fig:mpi-unphy}
\end{figure}

We also show a comparison of the fit to our calculation with the LQCD calculation of pressure and shear at $m_\pi=450$ MeV~\cite{Shanahan:2018nnv} in 
Fig.~\ref{fig:mpi-unphy}. 
In this case, the calculated total distributions also include gluon contributions from Eq.~(\ref{eq:convolution}).
The qualitative features of the lattice data are reasonably well reproduced, 
although at this mass the distribution of both pressure and shear reported in the lattice calculation are somewhat larger in the region beyond 1 fm than those found in the model. Further work will be needed to determine whether this difference arises from the input quark and gluon distributions in the pion or the nucleon calculations themselves.

For comparison with the LQCD result of Ref.~\cite{Shanahan:2018nnv}, we have also calculated the so-called ``mechanical radius'', which is defined as 
\be
\langle r^2_{\rm mech} \rangle = \dfrac{\int r^2 Z(r) d^3 r}{\int Z(r) d^3 r} ,
\ee
where $Z(r) = \frac{2}{3} s(r) + p(r)$. 
The LQCD result is $0.51\ {\rm fm}^2$ using the z-expansion and $0.57\  {\rm fm}^2$ for a tripole representation of $D(t)$ at $m_{\pi} = 450\ {\rm MeV}$. 
We find $\langle r^2_{\rm mech} \rangle = 0.42\ {\rm fm}^2$ at the mass used in the LQCD calculations.

\section{Conclusion}

In summary, we have calculated the pressure distribution and shear inside the proton using the cloudy bag model, which restores chiral symmetry to the MIT bag model. 
Compared with the experimental result for the quark contribution, the pion cloud, which peaks at the bag surface in this simple model, accounts reasonably well for the nonzero distributions at larger radii.
Given that the LQCD results correspond to a much larger scale, it seems most appropriate to compare the total distributions with our model results. While the qualitative features are reproduced at $m_\pi=450$ MeV, there is some discrepancy for radii beyond $0.8\ {\rm fm}$. It will be fascinating to see whether this discrepancy persists in future lattice calculations. On the other hand, at the physical quark mass, the comparison of the quark contribution to the pressure and shear inside the proton with the chiral model reveals quite satisfactory agreement. The pion contribution dominates for radii larger than 0.8 fm.

The model calculations presented here can be improved in future work. Rather than using a static (infinitely heavy) nucleon one can generate approximate eigenstates of momentum in various ways. These will have the effect of smoothing out the discontinuity at the bag radius. However, such improvements will not change the qualitative features reported here, namely a positive pressure distribution in the interior of the nucleon associated primarily with confined quarks and a negative pressure distribution outside generated by the pion cloud.

We look forward to the new experiments at JLab and the future 
Electron-Ion Collider (EIC)~\cite{Accardi:2012qut, AbdulKhalek:2021gbh}, which will have much extended kinematical coverage and should provide more precise information on the gravitational form factors of the nucleon.

\section*{Declaration of competing interest}
The authors declare that they have no known competing financial interests or personal relationships that could have 
appeared to influence the work reported in this paper.

\section*{Acknowledgements}
We would like to thank Phiala Shanahan for helpful discussions.
This work was supported by the University of Adelaide and the Australian Research Council through the Centre of Excellence for Dark Matter Particle Physics (CE200100008) and Discovery Project DP180100497.



\begin{thebibliography}{99}
\bibitem{dpiBfit}
$D_{\pi /B}$ is well fit up to 5 GeV$^2$ by the form $D_{\pi/B}\,(t) = A\, e^{-D(-t)}+B\, e^{-E(-t)}+C$, 
with $(A,B,C,D,E)=(0.23,0.0467,-0.0548,0.0932,2.02)$ and $(0.16,0.0187,-0.0449,0.0826,43.3)$ for $\pi \, N$ and $\pi \, \Delta$, respectively at $m_\pi=140$ MeV and $(0.10,-0.013,-0.0354,0.0702,1.46)$ and $(0.59,-0.509,-0.0298,0.0811,0.08)$ at $m_\pi=450$ MeV. Here, $A$, $B$, and $C$ are dimensionless, and $D$ and $E$ have units of fm$^2$.
\bibitem{psfits}
The constants $(A,B,C,D,E)$ are $(1.12,-1.67,5.15,2.11,4.92)$ and  $(2.40,-1.73,5.23,5.19,5.15)$  at $m_\pi$ equal to 140 and 450 MeV, respectively. Here, $A$ has units of GeV fm$^{-3}$, $D$ has units of fm$^{-4}$, and $B$, $C$, and $E$ have units of fm$^{-1}$.
\end{thebibliography}
\end{document}